\documentclass[conference]{IEEEtran}
\IEEEoverridecommandlockouts

\usepackage{placeins}
\usepackage{multicol}
\usepackage{flushend}
\usepackage{amsmath,amssymb,amsfonts}
\usepackage{array}
\usepackage{graphicx}
\usepackage{textcomp}
\usepackage{xcolor}
\usepackage{url}
\usepackage{lipsum}
\usepackage{wrapfig}
\usepackage{float}
\usepackage{hyperref}
\usepackage{orcidlink}
\usepackage{subcaption}
\usepackage{fancyhdr,lipsum}
\usepackage[pscoord]{eso-pic}
\usepackage{amsmath, amssymb}
\usepackage{algorithm}
\usepackage{algpseudocode}

\usepackage[utf8]{inputenc}
\usepackage{csquotes}
\usepackage[english]{babel}
\def\BibTeX{{\rm B\kern-.05em{\sc i\kern-.025em b}\kern-.08em
    T\kern-.1667em\lower.7ex\hbox{E}\kern-.125emX}}

\begin{document}

%%%%%%%%%% Choose the appropriate title for your group size %%%%%%%%%%
%%%%%%%%%%%%%%%%%%%%%%% Delete the rest %%%%%%%%%%%%%%%%%%%%%%%%%%%%%%

\title{
Federated Learning for Diabetic Retinopathy Diagnosis: Enhancing Accuracy and Generalizability in Under-Resourced Regions

}

\author{\IEEEauthorblockN{1\textsuperscript{st} Gajan Mohan Raj \orcidlink{0009-0001-2225-9756}}
\IEEEauthorblockA{\textit{High Technology High School}\\
Manalapan, USA \\
gajanmohanraj@gmail.com}
\and
\IEEEauthorblockN{2\textsuperscript{nd} Michael G. Morley \orcidlink{0000-0001-6373-7008}}
\IEEEauthorblockA{\textit{Massachusetts Eye and Ear} \\
\textit{Harvard Medical School}\\
Boston, USA \\
michael\_morley@meei.harvard.edu}
\and
\IEEEauthorblockN{3\textsuperscript{rd} Mohammad Eslami \orcidlink{0000-0002-6570-2826}}
\IEEEauthorblockA{\textit{Harvard Ophthalmology AI Lab} \\
\textit{Harvard Medical School}\\
Boston, USA \\
moha\_esla@meei.harvard.edu}
}

\maketitle
% \noindent\fbox{%
%     \parbox{\textwidth}{%
%         \begin{center}
%         *Corresponding author
%         \end{center}
%     }%
% }
% Citation Command Library
% \cite{niddk_diabetes}
% \cite{ting_meta_analysis}
% \cite{rajalakshmi_ai_dr_detection}
% \cite{thomas_dl_dr_screening}
% \cite{bhuiyan_quality_of_care}
% \cite{abadi_federated_learning}
% \cite{loggins_ai_healthcare}
% \cite{williams_data_mining}
% \cite{kuner_gdpr}
% \cite{smith_privacy_preserving}
% \cite{mcmahan_communication_efficient}
% \cite{zhou2021deep}

\newcommand{\placetextbox}[3]{
 \setbox0=\hbox{#3}
 \AddToShipoutPictureFG*{ \put(\LenToUnit{#1\paperwidth},\LenToUnit{#2\paperheight}){\vtop{{\null}\makebox[0pt][c]{#3}}}
 }
 }
\placetextbox{.23}{0.055}{\small{978-1-6654-7345-3/22/\$31.00 ~\copyright2024 IEEE}}

\begin{abstract}
Diabetic retinopathy is the leading cause of vision loss in working-age adults worldwide, yet under-resourced regions lack ophthalmologists. Current state-of-the-art deep-learning systems struggle at these institutions due to limited generalizability. This paper explores a novel federated learning system for diabetic retinopathy diagnosis with the EfficientNetB0 architecture to leverage fundus data from multiple institutions to improve diagnostic generalizability at under-resourced hospitals while preserving patient-privacy. The federated model achieved 93.21\% accuracy in five-category classification on an unseen dataset and 91.05\% on lower-quality images from a simulated under-resourced institution. The model was deployed onto two apps for quick and accurate diagnosis.
\end{abstract}

\section{Introduction}

%Diabetes mellitus is a healthcare condition characterized by abnormally high glucose levels in the bloodstream due to insufficient insulin production or inability to use insulin effectively \cite{niddk_diabetes}. Diabetic retinopathy (DR) is a common vision-threatening complication of diabetes, that affects blood vessels in the retina. If left untreated properly, diabetic retinopathy can lead to severe vision loss and blindness. It is estimated that 103 million individuals worldwide are affected by diabetic retinopathy, and this prevalence is expected to grow to 161 million individuals worldwide by 2045. Recent projections to 2030 also estimate that rates of increase in DR prevalence in under-resourced regions such as the Middle East and Africa range from 20.6\% to 47.2\% which is much higher than rates of increase in high-income regions such as the United States \cite{ting_meta_analysis}.

Diabetes mellitus often leads to diabetic retinopathy (DR), a condition that damages retinal blood vessels and can cause severe vision loss or blindness. Currently, 103 million people worldwide are affected by DR, with this number expected to rise to 161 million by 2045. Projections indicate that DR prevalence in under-resourced regions like the Middle East and Africa could increase by 20.6\% to 47.2\% by 2030, outpacing growth in high-income areas like the United States \cite{wong2023diabetic}.

\subsection{Motivations}

\subsubsection{Ophthalmologists Shortage}
Early detection of DR is crucial to prevent vision loss, yet many regions worldwide face a shortage of ophthalmologists required for diagnosis. In Sub-Saharan Africa countries, there is an average of approximately 2.5 ophthalmologists per a million population, compared with a ratio of 56.8 ophthalmologists per a million population in the United States \cite{resnikoff2020ophthalmologists, feng2020national}. This limited access to ophthalmologists leads to delayed diagnosis and increased risk of blindness for people in these under-resourced regions. Thus, there is a need for alternative diagnosis methods.

%Advances in Artificial Intelligence, particularly with deep learning, have led to innovations in DR diagnosis. Using deep learning models for DR diagnosis can allow non-ophthalmologist clinicians located in under-resourced regions to diagnose diabetic retinopathy more accurately. 
%However, for a deep learning-based diagnostic system to be applicable in real-world scenarios, it must achieve high diagnostic accuracy across a broad population with diverse demographic characteristics, while also considering regional variations in diagnostic practices and quality. 
%Deep learning models trained on a specific medical institution’s dataset have been shown to overfit at the institution’s data while being unable to generalize effectively to data at other medical institutions. For example, Google developed a deep learning system for DR detection in 2020, but it failed in real-life testing due to an inability to generalize at other medical institutions and a 60-90 second screening time despite a proposed accuracy of 90\% \cite{abadi_federated_learning}.
\subsubsection{Current Challenges of AI in DR Diagnosis}
While advances in AI have improved DR diagnosis, current AI-diagnostic systems struggle to be effective in real-world settings, due to over-fitting to a specific institution's data and poor generalization across diverse populations. For instance, Google's 2020 DR detection model, despite its 90\% proposed accuracy, failed in real-world testing due to poor generalization and long screening times \cite{beede2020human}.

\subsubsection{Resource Constraints in DR Imaging}
In addition to having a lack of trained ophthalmologists, many under-resourced healthcare institutions are located in sparse, remote regions. The retinal images collected in under-resourced regions are also of lower quality due to a lack of adequate imaging devices. Deep learning algorithms require large datasets of high-quality images to perform reliably and accurately. The inadequacy in both the amount and quality of retinal images at medical institutions of under-resourced regions makes it difficult for these institutions to build accurate DR-diagnosis deep learning systems. Thus, it is crucial for medical institutions in under-resourced regions to have access to high-quality DR data to improve their diagnosis models.

\subsection{Solutions}
%One solution could be to collect fundus data from different medical institutions found worldwide. A centralized deep learning model for DR diagnosis can be trained on this accumulated dataset situated on a central server, as can be seen in Fig. 1A. This solution would address the challenges of data inadequacy, lack of data diversity, and data quality. However, this centralized data-sharing approach faces restrictions due to patient-privacy, intellectual property, and data ownership concerns. For example, the General Data Protection Regulation in Europe and United States Health Insurance Portability and Accountability Act (HIPAA) both restrict the sharing of patient data across institutions \cite{williams_data_mining}, \cite{kuner_gdpr}. Patients may also have privacy and confidentiality concerns which makes it difficult for private medical data to be shared out of the originating institution. Therefore, it is difficult to implement a centralized data-sharing approach for DR diagnosis on the international scale.

One approach is to gather fundus data from various global medical institutions and train a centralized DR diagnosis model on this combined dataset (Fig. 1A). However, centralized data-sharing faces significant challenges due to patient privacy, intellectual property, and data ownership concerns. Regulations like Europe's GDPR and the U.S.'s HIPAA \cite{annas2003hipaa, book} restrict sharing patient data across institutions, making international implementation of a centralized data-sharing model difficult.

Federated learning (FL) enables collaborative, decentralized training of models across multiple clients without transferring data between them. Instead of gathering data from various locations, FL trains a shared model on a central server while keeping the data at its origin by learning from CNN architectures. This approach allows medical institutions to contribute to AI-driven systems for diagnosis, treatment, and disease monitoring without compromising patient privacy \cite{smith_privacy_preserving}. This approach can be valuable in regions like Sub-Saharan Africa, where quality data and trained ophthalmologists are scarce. 
In this work, we explore the use of federated learning for DR diagnosis and severity assessment, demonstrating with experiments how federated learning can also be a solution for under-resourced regions.

\begin{figure}[t]
    \centering
    \hspace*{-0.5cm}
\includegraphics[width=0.3\textwidth]{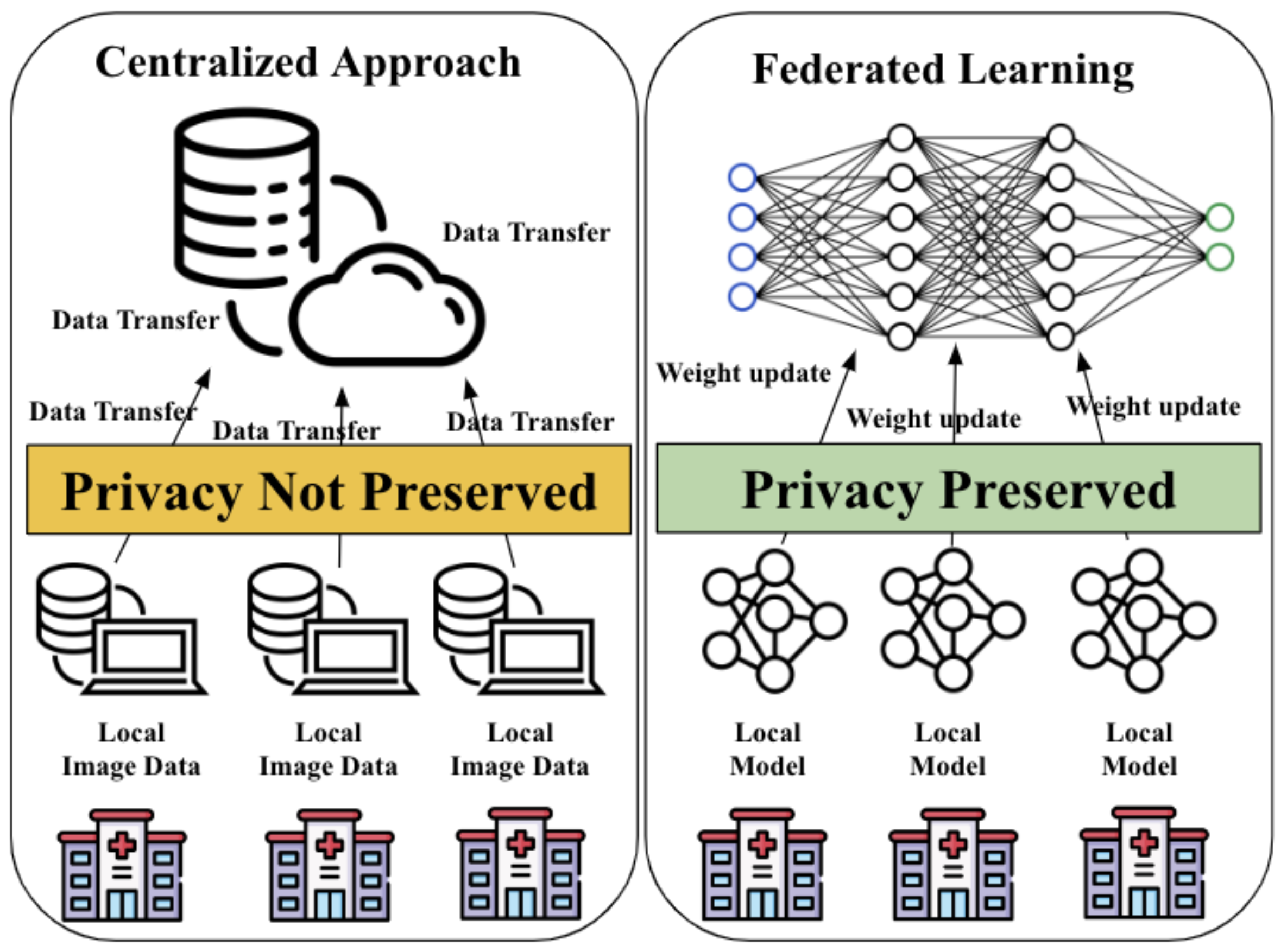}
    \caption{Centralized data-sharing system (a) vs Federated learning-driven system (b)
} 
    \label{fig:6}
\end{figure}

\section{Materials and Methodology}
\subsection{Background}

\subsubsection{Federated learning-driven system framework} Patient privacy, confidentiality, and data ownership concerns prevent the development of a centralized data-sharing system in which data would be shared to a central server (depicted in Fig. 1A). Therefore, a federated learning-driven system is proposed. Fig. 1 depicts the federated learning system as compared to the centralized data-sharing system. As depicted in Fig. 1B, the medical institutions send updates of model parameters to the central server after training a local copy of a diagnosis model on local data. The central server compiles the updates for the federated model and distributes the revised model parameters to all participating institutions for continued training or deployment. As can be observed through this process, patient privacy, confidentiality, and data-ownership concerns are significantly protected in the federated learning-driven system since raw image data is not shared.

The federated learning-driven system implemented in this work uses the FederatedAveraging (FedAvg) algorithm (Fig. 2).

\begin{figure}[H]  % Start of figure environment
\hspace*{-0.5cm}
\small  % Start of smaller font size
\centering  % Center the content

\noindent\textbf{FederatedAveraging Algorithm.} The $K$ institutions are indexed by $k$, $B$ is the size of local minibatch $b$, $E$ is the number of local epochs in each round of institution update, $\eta$ is the learning rate, and $\ell(\omega, b)$ is the local loss function.

\vspace{0.2cm}

\hrule
\vspace{0.2cm}

\noindent\textbf{Central Server executes:}
\begin{itemize}
    \item Initialize $\omega_0$
    \item For each round $t = 1, 2, \dots$ do
    \begin{itemize}
        \item For each institution $k = 1, 2, \dots, K$ in parallel do
        \begin{itemize}
            \item $\omega_{t+1}^k \leftarrow \text{InstitutionUpdate}(k, \omega_t)$
        \end{itemize}
        \item $\omega_{t+1} \leftarrow \sum_{k=1}^{K} \frac{n_k}{N} \omega_{t+1}^k$
    \end{itemize}
\end{itemize}

\vspace{0.2cm}

\noindent\textbf{InstitutionUpdate}$(k, \omega)$:
\begin{itemize}
    \item For each local epoch $i$ from 1 to $E$ do
    \begin{itemize}
        \item For each minibatch $b$ do
        \begin{itemize}
            \item $\omega \leftarrow \omega - \eta \nabla \ell(\omega; b)$
        \end{itemize}
    \end{itemize}
    \item Return $\omega$ to central server
\end{itemize}

\normalsize  % Return to normal font size

\caption{FederatedAveraging Algorithm \cite{mcmahan_communication_efficient}}
\label{fig:federated_averaging}
\end{figure}

The process begins with the central server initializing the model parameters \(\omega_0\), which are then distributed to each participating institution. Training proceeds in multiple rounds, indexed by \(t\). In each round, the central server sends the current global model \(\omega_t\) to all \(K\) institutions. Each institution \(k\) then executes the \texttt{InstitutionUpdate} process, wherein it trains the model locally using its data for \(E\) epochs. During each epoch, the model parameters \(\omega\) are updated using the SGD optimizer to minimize the local loss function \(\ell(\omega, b)\) for each minibatch \(b\).

After local training, institutions send their updated models \(\omega_{t+1}^k\) to a central server, which aggregates them into a new global model \(\omega_{t+1}\) by computing a weighted average based on the sizes of the local datasets \(n_k\). This process repeats across multiple rounds \(t\) until model convergence. The FedAvg algorithm was implemented using TensorFlow Federated (TFF). 

\subsubsection{Convolutional Neural Networks (CNNs)}
CNNs are a type of neural network, often utilized for image classification and computer vision. Local CNNs located at each medical institution serve as the basis for the federated learning-driven system. 
The four CNN architectures included in our study are EfficientNetB0, MobileNetV2, InceptionResnetV2, and Xception. These four CNN architectures were pre-trained on the ImageNet dataset which contains over 14 million labeled images. 

\subsection{Dataset}
Three publicly available datasets from different regions of the world were used to develop and evaluate our framework: the DDR dataset (12,256 fundus photos, China)\cite{herrero_ddr_dataset}, the EyePACS dataset (11,673 fundus photos, California, USA)\cite{kaggle_diabetic_retinopathy_competition}, and the APTOS dataset (3,357 images, India)\cite{kaggle_aptos2019_blindness_detection}. 
The fundus images were classified into five categories of DR severity: No DR, Mild DR, Moderate DR, Severe DR, and Proliferative DR.
We will treat each of these datasets as a separate medical institution.

\subsection{Implementation Details}

\subsubsection{Preliminary Study (Finding Best CNN Architecture)}
In order to ensure that the federated learning-driven system is accurate while also being resource-efficient, we  considered which CNN architecture to utilize as the architecture for the single-institution model. For this reason, the four different CNN architectures mentioned earlier were trained, validated, and evaluated on the DDR dataset. 

Prior to training and testing, each CNN architecture was modified by adding a batch normalization layer, a dense layer, a dropout layer (rate of 0.45), and a final dense layer with the softmax activation function. 
The modified diagnostic models were then trained, validated, and tested on the DDR dataset to accurately classify each image into one of the five severity levels. 
We used five-fold cross-validation to evaluate classification performance. 
For each fold, the model ran through forty epochs. 
An early stopping algorithm of 10 epochs was also implemented to prevent over-fitting.
We used the stochastic gradient descent (SGD) optimizer with a learning rate of 0.001. 
The performance of each of the CNN architectures was assessed with the following metrics: ROC AUC score, Accuracy, and Model Size in megabytes (MB). 
The CNN architecture that performed the best in all three of these metrics (high ROC AUC Score and accuracy; low model size) was chosen as the architecture for the local CNN-based diagnostic systems at individual institutions.

\subsubsection{Federation of Institutions' Diagnosis Models}

To evaluate the performance of a federated learning-driven system for DR diagnosis, a simulation of 3 medical institutions was designed. In this work, we simulated a scenario of two medical institutions, H1 and H2, being well-resourced institutions and one medical institution, H3, being an under-resourced medical institution.

%Each medical institution was simulated by using a different dataset of fundus images and their DR label. 
The DDR dataset was used as H1, the EyePACS dataset was used as H2, and the APTOS dataset was used as H3. 
In order to simulate H3 as an under-resourced institution, the images of the APTOS dataset were purposefully reduced to lower quality by adjusting the jpeg image quality to a random value between 30 and 50. To prepare for Experiment 1, which is discussed in detail later, we set aside 3,250 images each from H1 and H2, creating an independent test set of 6,500 high-quality images. Then, 10\% of images were set aside from the remaining fundus images in H1, H2, and H3 to be used as each institution's test set. These test sets were used in Experiment 2, which is also discussed in detail later. The remaining 90\% of each dataset was used for training and validation. 
All three datasets were imbalanced, so random data augmentation was performed for oversampling. Horizontal flips with 50\% probability, 30° random rotations with 25\% probability, random brightness contrasts with 25\% probability, and Gaussian filter with 100\% probability were the four data augmentation techniques performed. 

Two experiments were performed to evaluate the federated learning-driven system's performance compared to the three single-institution systems' performance. 
First in \textit{experiment 1}, the ability of the federated learning-driven system to improve accuracy by utilizing data from multiple institutions was evaluated. Fig. 3A depicts this experiment. The local models at each of the three medical institutions, H1, H2, and H3, were trained independently, and their gradients were communicated to a central server to update the federated model. This process continued until the federated model converged. 
This federated model and the three local models were then tested on the independent test set of 6500 high-quality fundus images. Accuracy was used as the primary metric of evaluation.
 
Then, in \textit{experiment 2}, the generalizability of the federated learning-driven system to improve performance on lower-quality fundus images, compared to local models, was evaluated. The federated learning framework was setup, similar to the first experiment, but each local model was tested on its own institution's test set in addition to the test set of the other institutions. 
The federated model was also tested on the test set of each of the three institutions. Fig. 3B displays this experiment. Accuracy was used as the primary metric of evaluation once again.

\begin{figure}[H]
    \centering
    \hspace*{-0.5cm}
    \begin{subfigure}[b]{0.45\textwidth}
        \centering
        \includegraphics[width=\textwidth]{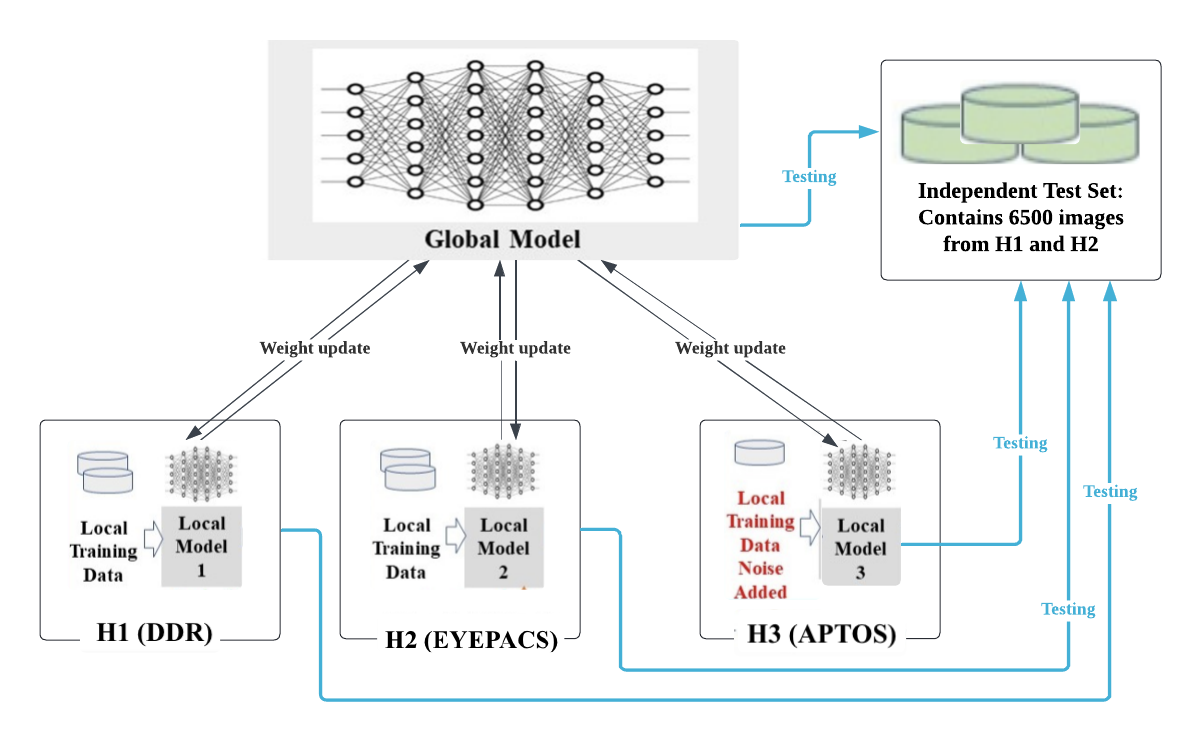}
        \caption{Experiment 1}
        \label{fig:3}
    \end{subfigure}
    \hspace*{0.02cm}
    \begin{subfigure}[b]{0.45\textwidth}
        \centering
        \includegraphics[width=\textwidth]{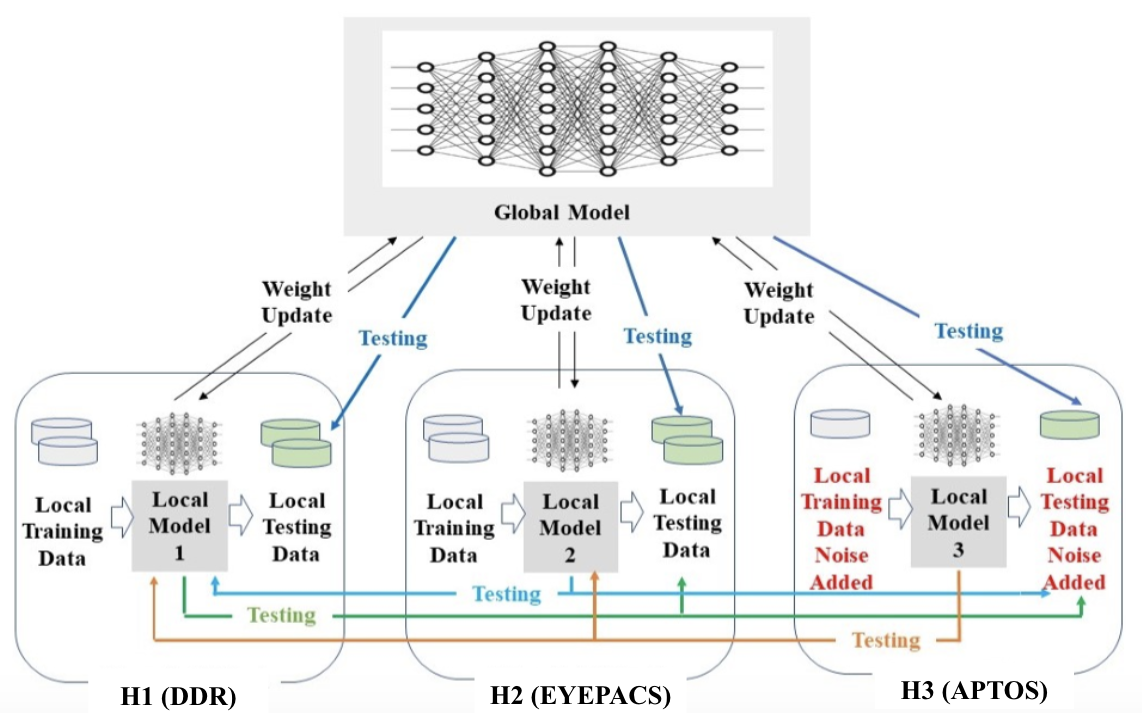}
        \caption{Experiment 2}
        \label{fig:2B}
    \end{subfigure}
    \caption{Comparison of federated experiments: Experiment 1 (a) and Experiment 2 (b)}
    \label{fig:3}
\end{figure}

\begin{figure*}[tbp]
    \centering
    \hspace*{-0.5cm}
    \begin{subfigure}[b]{0.4\textwidth}
        \centering
        \includegraphics[width=\textwidth]{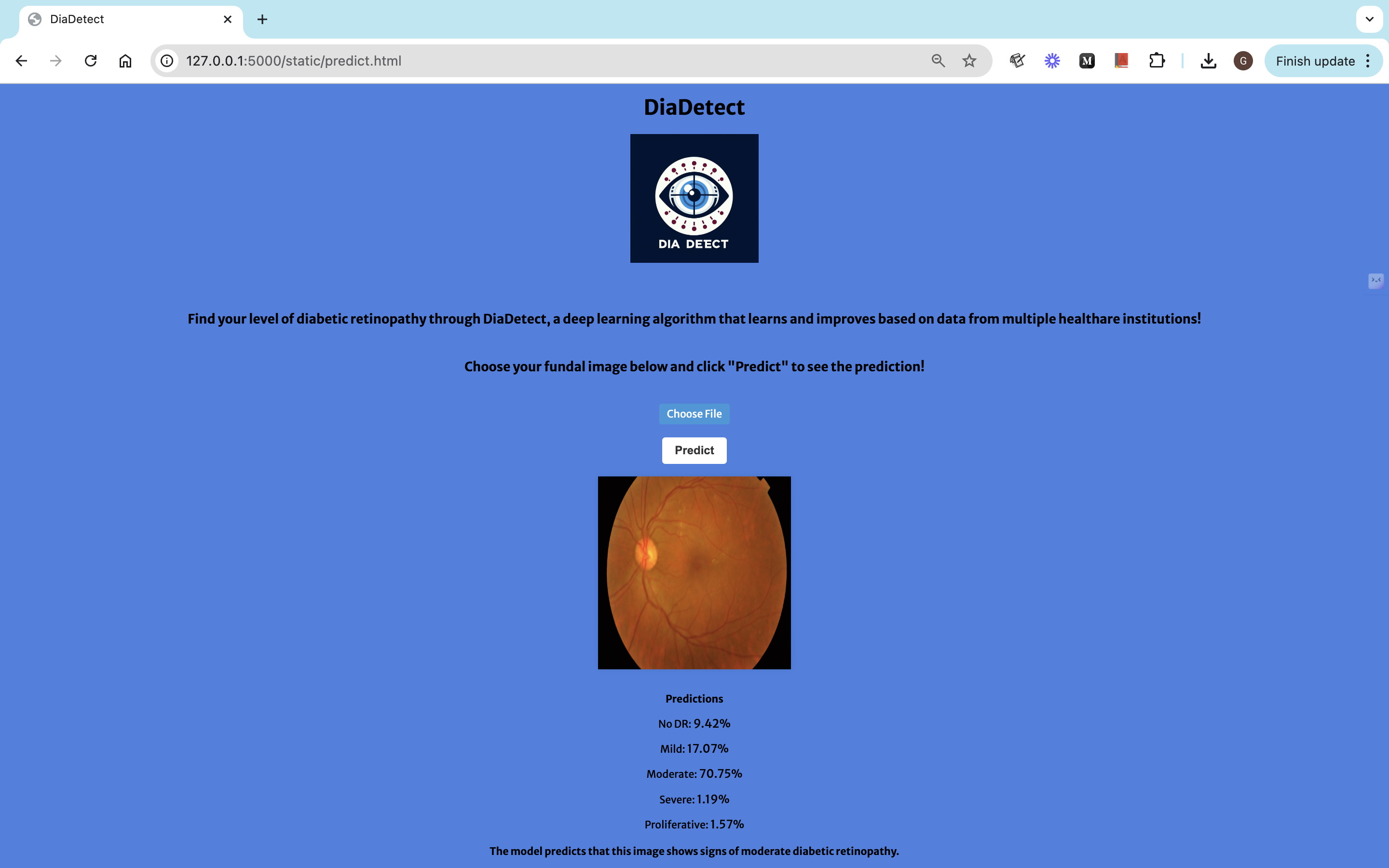}
        \caption{Web Application}
        \label{fig:2A}
    \end{subfigure}
    \hspace*{0.02cm}
    \begin{subfigure}[b]{0.55\textwidth}
        \centering
        \includegraphics[width=\textwidth]{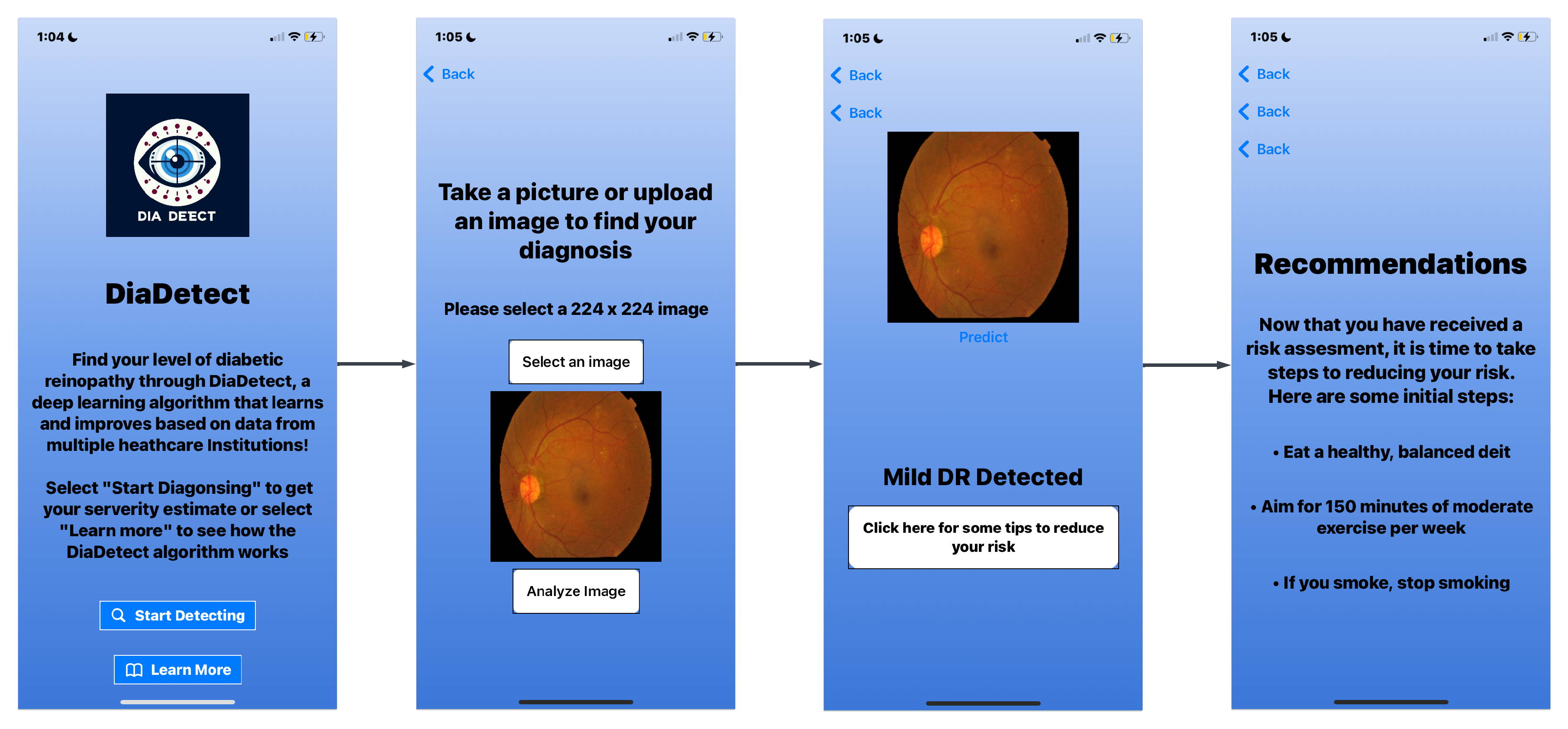}
        \caption{iOS Mobile App}
        \label{fig:2B}
    \end{subfigure}
    \caption{Developed demonstrations and applications.}
    \label{fig:2}
\end{figure*}

\subsection{Prototype Demonstration}

As a prototype demonstration of how the developed AI model could be used by clinicians to address the shortage of ophthalmologists, a web application and an iOS mobile application for DR diagnosis were developed (Fig. 4). 

After the two evaluation experiments explained above, the federated model was saved as a .h5 file. In the development of the web application, HTML and CSS were used for the front-end while Flask was used to host the federated model. In the development of the iOS mobile application, the CoreML framework and the Swift programming language were used. CoreML was used to integrate the downloaded federated model within the application while Swift was used to develop the interface of the application. Both applications allow users to submit a fundus image to get an accurate multi-class diagnosis of DR.

\section{Results}

\subsection{Preliminary Study}

The four CNN architectures, EfficientNetB0, MobileNetV2, InceptionResnetV2, and Xception, were trained and tested using five-fold cross-validation on the DDR dataset. As shown in Table 1, the EfficientNetB0 architecture outperformed the other three architectures with an accuracy of 0.790 and ROC AUC Score of 0.934. It also had the second-lowest model size of 29 MB. Due to its balance of accuracy and resource-efficiency, the EfficientNetB0 architecture was chosen as the CNN architecture located at each local institution.

 \begin{table}[!h]
\centering
\caption{Model Comparison For Preliminary Testing: Accuracy, ROC AUC Score, and Model Size}
\begin{tabular}{|l|c|c|c|}
\hline
\textbf{Architecture} & \textbf{Accuracy} & \textbf{ROC AUC} & \textbf{Model Size} \\ \hline
MobileNetV2       & 0.671 & 0.866 & 14 MB  \\ \hline
InceptionResnetV2 & 0.728 & 0.929 & 215 MB \\ \hline
Xception          & 0.775 & 0.912 & 88 MB  \\ \hline
 EfficientNetB0  & 0.790 & 0.934 & 29 MB  \\ \hline
\end{tabular}
\label{tab:model_comparison}
\end{table}

 \subsection{Experiment 1: Accuracy Evaluation}

 The three local models at each medical institution and the federated model were evaluated for accuracy on an independent test set of 6500 images after the federated learning training and validation process. As shown in Table 2, the federated model outperformed the local models on the test-set with an accuracy of 0.9321 as compared to 0.8876, 0.9036, and 0.7148 from H1 Model, H2 Model, and H3 Model. Fig. 5 further demonstrates the performance of the federated model across all five DR classifications on the test-set. These results suggest that the federated model is able to integrate data from multiple medical institutions to improve DR diagnosis.

\begin{table}[!h]
\centering
\caption{Model Accuracy Comparison on Independent Test Set}
\begin{tabular}{|l|c|}
\hline
 \textbf{Model} & \textbf{Accuracy} \\ \hline
H1 Model (DDR) & 0.8876 \\ \hline
H2 Model (EYEPACS) & 0.9036 \\ \hline
H3 Model (APTOS) & 0.7148 \\ \hline
Federated model & 0.9321 \\ \hline
\end{tabular}
\label{tab:model_accuracy_comparison}
\end{table}

\begin{figure}[h]
    \centering
    \hspace*{-0.5cm}
\includegraphics[width=0.37\textwidth]{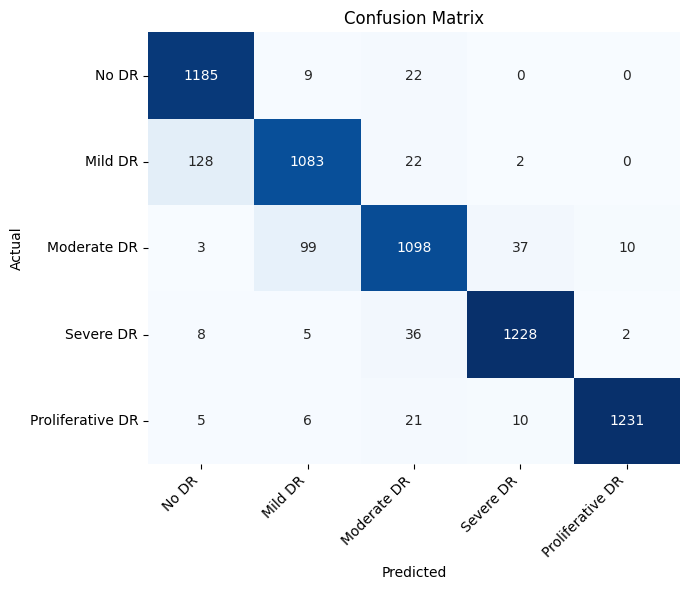}
    \caption{Confusion Matrix of Federated Model on Independent Test-set
} 
    \label{fig:5}
\end{figure}

\subsection{Experiment 2: Generalizability Evaluation}

There may exist discrepancies in imaging equipment, differences in patient population, and fundus image preparation in real-world scenarios. As a result, a generalizibility evaluation was simulated in this work by purposefully reducing the image quality of all fundus images at H3's dataset. This institution also consisted of the least amount of images, thus simulating the data-inadequacy of medical institutions in under-resourced regions. After the federated learning training process, each local model was tested on its own local test-set as well as the test-set of the other two medical institutions. The federated model was also tested on the test-sets of all three medical institutions. Fig. 6 displays the results of this experiment. 

 As can be seen in Fig. 6, the federated model performed best on the test-sets of all three medical institutions. The federated model achieved an accuracy of 0.9708, 0.9634, and 0.9105 at the test-sets of H1, H2, and H3. The local models did not perform as well on other test-sets in addition to their own test-sets. 

 \begin{figure}[!h]
    \centering
    \hspace*{-0.5cm}
\includegraphics[width=0.4\textwidth]{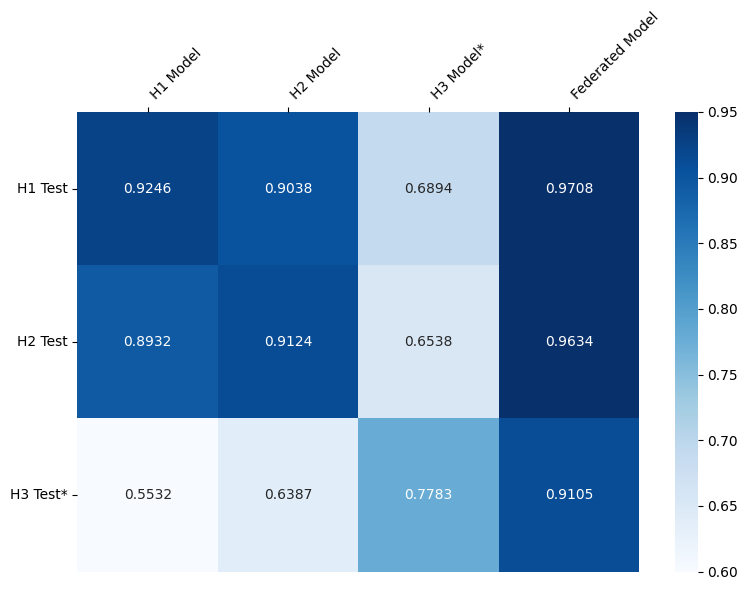}
    \caption{Performance of Each Model at Each Test-Set
} 
    \label{fig:6}
\end{figure}
 
\section{Discussion}
Medical institutions found in under-resourced regions require resource-efficient, fast deep learning models. 
After the CNN architecture evaluation of accuracy, ROC AUC score, and model size, the EfficientNetB0 model was chosen as the CNN architecture for the local models and federated model.
%due to its compact size of 29 MB, accuracy of 0.790, and high ROC AUC score of 0.934. These features, specifically the small model size, allows the federated learning-driven system to be deployed to under-resourced regions where computational resources may be limited. 
%To add on, deep learning models depend on an abundant amount of high-quality data to perform accurately. Our experiments demonstrated that as well. 
Investigating Table II and Fig. 6, we can comprehend the following outcomes:
\begin{itemize}
    \item Once we diluted the H3 dataset with low-quality photos, it was observed that the model of H3 performed poorly while the models of H1 and H2 performed accurately when tested on the independent test-set. 
This illustrates the dilemma that medical institutions found in under-resourced regions face. 
These institutions have an inadequate amount of high-quality data that are required to build effective deep learning models for DR diagnosis.  Federated learning provided an innovative solution to this problem as it was able to learn from the local models found at all three medical institutions to improve DR diagnosis accuracy to 0.9321.
%It achieved an accuracy of 0.9321, an increase of 4.45\%, 2.85\%, and 21.73\% compared to the performance of local models at H1, H2, and H3.
    \item It has previously been recognized that diagnostic models for DR diagnosis that have been trained on data from a particular institution may not perform as well on the data found at other institutions, as evidenced by Google's failure in developing a generalizable DR diagnosis model. This issue was recognized in this study as well; the local models at H1 and H2 performed well on their own test-sets and each other institution's test-sets.
    However, the local models of H1 and H2 performed poorly on the lower-quality data at H3 with accuracies of 0.5532 and 0.6387, respectively. These results demonstrate that although models trained on high-quality data can perform well on other test-sets containing high-quality data, they do not generalize as well to datasets with lower-quality data. 
    However, the federated model which learned from local models at all three institutions performed impressively even on the test-set at H3 with an accuracy of 0.9105. This may be because the federated model learned from diverse, high-quality data found at H1 and H2 while still incorporating the lower-quality data found at H3.     
\end{itemize}

\section{Conclusion}

Ultimately, the federated learning system’s high accuracy of 93.21\%, ability to generalize for under-resourced institutions, and resource-efficiency make it a promising solution for improving DR screening in hospitals of rural areas and developing countries where there is a shortage of trained ophthalmologists. 
The federated learning-driven system can address delayed diagnosis of DR and the risk of blindness in millions of people of under-resourced countries. 
The federated learning-driven system accurately classifies fundus images into five categories.
This precise classification helps clinicians recommend appropriate treatments for each patient. 
Additionally, the federated learning-driven system's ability to not use private medical data found at local institutions for training mitigates concerns for patient privacy, confidentiality, and data-ownership. 
Finally, the system can allow healthcare institutions worldwide to collaborate on developing accurate DR diagnosis models. However, other issues such as the development of more robust privacy preserving solutions, scalability across healthcare institutions, and optimizing client-server communication efficiency will still need to be explored in future work.

This work has demonstrated the ability of a federated learning-driven system to improve diagnostic model accuracy and generalizibility, even at under-resourced institutions with lower-quality data. All in all, this research provides a new collaborative framework for DR diagnosis which improves diagnostic accuracy and generalizability. 

\bibliographystyle{IEEEtran}
\bibliography{refs}

 %\\ \\ \\

%\newpage % This command starts a new page before the appendix section 
%\\ \\ \\ \\ \\ \\ \\\ \\\ \\\\\\\\\ \\ \\ \\ \\ \\ \\\ \\\ \\\\\\\\\ \\ \\ \\ \\ \\ \\\ \\\ \\\\\\\\\ \\ \\ \\ \\ \\ \\\ \\\ \\\\\\\\\ \\ \\ \\ \\ \\ \\\ \\\ \\\\\\\ \\ \\ \\ \\ \\ 

% Your main document content here

%\newpage % Start a new page before the appendix

%\onecolumn

%\newpage % Start a new page before the appendix

%\onecolumn

\par\leavevmode

\end{document}